\documentclass[conference]{IEEEtran}
\IEEEoverridecommandlockouts
\usepackage[table]{xcolor}
\usepackage{cite}
\usepackage{amsmath,amssymb,amsfonts}
\usepackage{algorithmic}
\usepackage{graphicx}
\usepackage{tcolorbox}
\usepackage{textcomp}
\usepackage{subfiles}
\usepackage{xcolor}

\usepackage{array}
\def\BibTeX{{\rm B\kern-.05em{\sc i\kern-.025em b}\kern-.08em
    T\kern-.1667em\lower.7ex\hbox{E}\kern-.125emX}}
\begin{document}

\title{Automated Code Review In Practice
}
\makeatletter
\newcommand{\linebreakand}{%
  \end{@IEEEauthorhalign}
  \hfill\mbox{}\par
  \mbox{}\hfill\begin{@IEEEauthorhalign}
}
\makeatother

\author{
    \IEEEauthorblockN{
        Umut Cihan\IEEEauthorrefmark{1}, 
        Vahid Haratian\IEEEauthorrefmark{1}, 
        Arda İçöz\IEEEauthorrefmark{1},\\
        Mert Kaan Gül\IEEEauthorrefmark{2}, 
        Ömercan Devran\IEEEauthorrefmark{2}, 
        Emircan Furkan Bayendur\IEEEauthorrefmark{2},\\
        Baykal Mehmet Uçar\IEEEauthorrefmark{2}, 
        Eray Tüzün\IEEEauthorrefmark{1}
    }
    \IEEEauthorblockA{
        \IEEEauthorrefmark{1}
        Bilkent University, Ankara, Türkiye\\
        umut.cihan@bilkent.edu.tr, vahid.haratian@bilkent.edu.tr
        arda.icoz@bilkent.edu.tr, eraytuzun@cs.bilkent.edu.tr
    }
    \IEEEauthorblockA{
        \IEEEauthorrefmark{2}
        Beko, İstanbul, Türkiye\\
        mertkaan.gul@beko.com, omercan.devran@beko.com
        emircanfurkan\_bayendur@beko.com, baykal.ucar@beko.com
    }
}

\maketitle

\maketitle

\begin{abstract}
\label{abstract}
Context: Code review is a widespread practice among practitioners to improve software quality and transfer knowledge. It is often perceived as time-consuming due to the need for manual effort and potential delays in the development process. Several AI-assisted code review tools (Qodo, GitHub Copilot, Coderabbit, etc.) provide automated code reviews using large language models (LLMs). The overall effects of such tools in the industry setting are yet to be examined.

Objective: This study examines the impact of LLM-based automated code review tools in an industry setting. 

Method: The study was conducted within an industrial software development environment that adopted an AI-assisted code review tool (based on open-source Qodo PR Agent). 238 practitioners across ten projects had access to the tool. We focused our analysis on three projects, encompassing 4,335 pull requests, of which 1,568 underwent automated reviews. Our data collection comprised three primary sources: (1) a quantitative analysis of pull request data, including comment labels indicating whether developers acted on the automated comments, (2) surveys sent to developers regarding their experience with the reviews on individual pull requests, and (3) a broader survey of 22 practitioners capturing their general opinions on automated code reviews.

Results: 73.8\% of automated code review comments were labeled as resolved. However, the overall average pull request closure duration increased from five hours 52 minutes to eight hours 20 minutes, with varying trends observed across different projects. According to survey responses, most practitioners observed a minor improvement in code quality as a result of automated code reviews.  

Conclusion: The LLM-based automated code review tool proved useful in software development, enhancing bug detection, increasing awareness of code quality, and promoting best practices. However, it also led to longer pull request closure times and introduced drawbacks such as faulty reviews, unnecessary corrections, and irrelevant comments. Based on these findings, we discussed how practitioners can more effectively utilize automated code review technologies.
\end{abstract}


\section{Introduction}
\newcommand{\companyName }{Beko}

\label{introduction}


Code review is a quality assurance process, with different adaptations across the industry, that involves developers checking each others' code changes  \cite{Baum2016_ClassificationSchemeForIndustrialCR}. Emerging as formal code inspections \cite{Fagan1976_codeInspections}, code review has evolved and become a lighter process often referred to as Modern Code Review (MCR) \cite{Davila2021_literatureReview}. MCR is characterized by being informal, tool-based, and regular \cite{Bacchelli2013_modernCodeReview}. The common benefits of the process are knowledge sharing, learning, defect identification, and code improvement \cite{Macleod2018_CRInTrenches, Sadowski2018_McrAtGoogle}.


To conduct a code review, developers spend time understanding other developers' work and where it stands in the overall project. The Open Source Software (OSS developers self-reported an average review time of 6.4 hours per week for code review \cite{Bosu2013_ImpactOfPeerCR}, whereas the Google case study reveals a lower number with 3.2 hours \cite{Sadowski2018_McrAtGoogle}. The numbers suggest that developers spend a considerable amount of time on reviews. 

Due to developers' workloads, developers often postpone code review tasks. In doing so, developers postpone the merge of code changes. In the same case study at Google, the median time for merge approval is under four hours, while for large change sets, it is five hours \cite{Sadowski2018_McrAtGoogle}. Higher median approval times are reported from various companies: 15.7 hours for Google Chrome, 20.8 hours for Android at Google, 17.5 hours for AMD, 14.7 hours for Bing, 19.8 hours for SQL, and 18.9 hours for Office at Microsoft\cite{Rigby2013_convergentContemporaryReview}.  This variation in approval times suggests differences across companies and projects. However, the actual costs of delays in critical code changes, which are not reflected in median or average statistics, could be more detrimental.  In a study of Microsoft developers, receiving timely feedback was cited as the top challenge regarding code review practices \cite{Macleod2018_CRInTrenches}.


Automation can potentially shorten the time for code approval and reduce the burden of manual effort on developers. In this regard, several attempts have been made to automate the code review process \cite{Hong2022_CommentFinder,Li2022_AutomatingCRByLargeScalePreTraining,Tufano2022_UsingPreTrainedModelsForCR,Li2022_AUGER,yu2024securitycodereviewlarge,fan2024_llmsForCR,vijayvergiya2024_aiAssistedAssesmentInModernCR,pornprasit2024_fineTuningPromptEng,nashaat2024_fineTuningLLMWithOrgDataForAutomatedCR,tang2024_codeAgentCR,Rasheed2024_AIPoweredCR}, focusing mainly on the assignment of reviewers \cite{Tufano2024_CodeReviewAutomation}. These efforts come from tools that generate code reviews \cite{Li2022_AUGER, Tufano2022_UsingPreTrainedModelsForCR, Li2022_AutomatingCRByLargeScalePreTraining}, predict the approval of code changes \cite{Shi2019_AutoCRByLearningRev, Li2019_DeepReview}, and fix code according to code reviews \cite{michele2019_NMT, Thongtanunam2022_AutoTransform}. In our study, we focus on the automation of code review generation.


To generate code reviews, developers spend time understanding the code change, looking for mistakes, and detecting performance bottlenecks or general deviations from coding standards. There have been several attempts with pre-trained models to provide code reviews\cite{Gupta2018_IntelligentCRUsingDL, Li2019_DeepReview, Shi2019_AutoCRByLearningRev}. With the introduction of ChatGPT\cite{ChatGPT}, there have been considerable automation efforts to automate code review generation \cite{davila2024_crInGenAI}. Several tools with code review capability were created, often working with OpenAI's LLMs such as GPT-4  \cite{CodeRabbit,Codium}. 


Automated code review tools are increasingly used in the industry \cite{davila2024_crInGenAI}. However, there is a lack of empirical evidence regarding their potential benefits. For example, the time saved through automated reviews could be offset by new problems they introduce. Additionally, the financial benefits of reducing developer effort may be negligible compared to the cost of operating such tools. 
To address this research gap, we conducted a study aimed at answering the following research questions:

\begin{itemize}
\item \textbf{RQ1:} How useful are LLM-based automated code reviews in the context of the software industry?

\item \textbf{RQ2:} How do LLM-based automated code reviews impact the pace of the pull request closure process? 

\item \textbf{RQ3:} How does the introduction of LLM-based automated code reviews influence the volume of human code review activities?

\item \textbf{RQ4:} How do developers perceive the LLM-based automated code review tools?
\end{itemize}


We collaborated with Beko, a multinational home appliances company, whose software division adopted an automatic code review tool based on the open-source Qodo (FormerlyCodiumAI) PR Agent \cite{Codium} using GPT-4 Turbo model \cite{Achiam2023_GPT4}. This tool provides automatic code review comments for each pull request across 10 projects and 22 project repositories. 


Our data collection process included three sources. First, we extracted data from the version control system and development platform Azure DevOps, which hosts the project repositories. This data encompassed pull request information, review comments, review comment labels, and comparisons between initial and final versions of pull requests. Developers were asked to label each review comment to indicate whether they had implemented the suggestions into the code. Second, the authors received a short survey for each pull request. Lastly, we conducted a broader survey involving 22 practitioners to gather their overall perceptions.



\section{Related Work}
\label{related_work}

Code reviews demand significant developer effort and time \cite{Macleod2018_CRInTrenches,Bosu2013_ImpactOfPeerCR}. These demands, along with the need for frequent context switching, have driven the push towards automation \cite{Tufano2021_TowardsAutomatingCR}. The automation in code review is a well-investigated part of the research, where the majority of efforts tackle the problem of reviewer assignment \cite{AlZubaidi2020_WorkloadAwareReviewerRecommendation,Asthana2019_WhoDoReviewerSuggestion,Jiang2019_WhoShouldMakeDecisionOnThisPR, Ying2016_EARec,Mirsaeedi2020_MitigatingTurnoverCRRecommendation,Ouni2016_SearchBasedRevRec, Rahman2016_Correct,Thongtanunam2015_WhoShouldReview,RSTRACE2021}. There is also considerable effort being put into other code review aspects. In 2018, Gupta et al.\cite{Gupta2018_IntelligentCRUsingDL} presented a model to match historical reviews with code snippets using supervised deep learning. In 2019, a Convolutional Neural Network (CNN) based model by Li et al. \cite{Li2019_DeepReview} and a framework utilizing CNN and LSTM by Shi et al. \cite{Shi2019_AutoCRByLearningRev} were presented to predict approvals of code changes. In 2022, Thontanunam et al. \cite{Thongtanunam2022_AutoTransform} presented a model to modify the source code automatically during code review processes.

In addition to the aforementioned efforts, several techniques were investigated for automation of code review generation, including information retrieval \cite{Hong2022_CommentFinder}, pre-trained models \cite{Li2022_AUGER,Tufano2022_UsingPreTrainedModelsForCR,Li2022_AutomatingCRByLargeScalePreTraining}, LLMs \cite{yu2024securitycodereviewlarge, fan2024_llmsForCR, vijayvergiya2024_aiAssistedAssesmentInModernCR}, LLM prompt fine-tuning \cite{pornprasit2024_fineTuningPromptEng}, fine-tuning LLMs with human feedback \cite{nashaat2024_fineTuningLLMWithOrgDataForAutomatedCR}, and LLM agents \cite{tang2024_codeAgentCR,Rasheed2024_AIPoweredCR}. The effectiveness of state-of-the-art code review automation has been investigated by Tufano et al.\cite{Tufano2024_CodeReviewAutomation}. In their study, Tufano et al. investigated their model based on a pre-trained Text-To-Text Transfer Transformer (T5) model \cite{Tufano2022_UsingPreTrainedModelsForCR}, \cite{Raffel2019_T5}, CommentFinder model using information retrieval techniques to recommend code reviews \cite{Hong2022_CommentFinder}, CodeReview Model using pre-training techniques \cite{Li2022_AutomatingCRByLargeScalePreTraining}, as well as ChatGPT \cite{ChatGPT} without specifying the version they used. They found that ChatGPT could serve as a competitive baseline for improving code, both in direct code-to-code transformations and in generating code based on comments (comment-to-code). In contrast, ChatGPT did not outperform the state-of-the-art in comment generation (code-to-comment). 

Usage of LLMs and generative artificial intelligence for code review generation has attracted other studies.  Davila et al. \cite{davila2024_crInGenAI} conducted a grey literature review regarding the usage of generative artificial intelligence for code reviews and showed how LLM-based tools like ChatGPT have been explored for code review.  Fan et al. \cite{fan2024_llmsForCR} explored the capabilities of LLMs on three code change-related tasks: code review generation, commit message generation and just-in-time comment update. They concluded that LLMs are promising for the tasks mentioned earlier. Watanabe et al. \cite{watanabe2024_useOfChatGPTForCR} investigated 229 review comments from 179 GitHub projects that included ChatGPT conversation links. Their analysis revealed that 30.7\% of the reactions to ChatGPT's answer were negative, with developers often citing the lack of extra benefits. In 2024, Vijayvergiya et al.\cite{vijayvergiya2024_aiAssistedAssesmentInModernCR} from Google and the University of Washington demonstrated their findings on the development and large-scale industry implementation of AutoCommenter. AutoCommenter is an LLM-backed automated code review system for four programming languages (C++, Java, Python, and Go). Their findings suggest the feasibility of developing an end-to-end automated code review system while achieving high end-user acceptance.

In this study, we aim to address the lack of longitudinal research on the impact of automated code review tools in software development. Unlike previous studies, our research examines the effects of a commercial LLM-based automated code review tool \cite{Codium} in real-world industry settings regarding its impact on development artifacts and developer perceptions. This study aims to provide practitioners with valuable insights into whether and how to adopt an LLM-based automated code review tool.

\section{Research Settings}
\newcommand{\surveyRespondents }{22}
\newcommand{\totalAnalyzedPRAfterCodium }{1568}
\newcommand{\totalAnalyzedPR}{4335}
\newcommand{\ProjectOne}{Project \#1} 
\newcommand{\ProjectTwo}{Project \#3} 
\newcommand{\ProjectThree}{Project \#2} 
\newcommand{\companyName}{Beko}
\label{Research Settings}
In this study, we conducted an evaluative case study to assess the impact of LLM-based automated code reviews in software development. We evaluated their effectiveness, influence on the speed of pull request closures, and changes in the volume of human code reviews. This section presents our research settings. Section \ref{Study-Object} introduces the study object. Section \ref{Study-Goal-and-Research-Questions} presents this study's goal and research questions. Our study involves both quantitative and qualitative data sources. Section \ref{Data-Collection-From-Azure-DevOps} describes how we collected quantitative data from the projects' repositories. Section \ref{Data-Collection-From-Surveys-and-Interviews} describes our qualitative data collection through surveys. Section \ref{Approach-Towards-Research-Questions} describes our approach toward research questions.
\subsection{Study Object}
\label{Study-Object}
\begin{figure}
    \centering
    \includegraphics[width=1\linewidth]{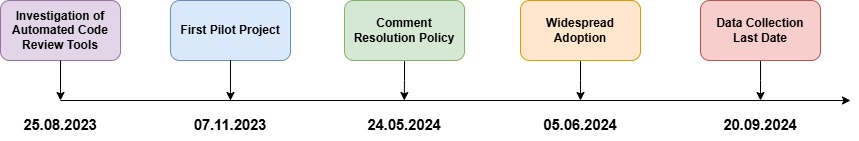}
    \caption{Data Collection Timeline}
    \label{fig:data-collection-timeline}
\end{figure}
We conducted our study within the software development division of \companyName{}, a multinational company. It operates in the consumer durables and electronics sectors. The software development division of \companyName{} is responsible for developing customer-facing and employee-facing software with 238 practitioners. 


Figure \ref{fig:data-collection-timeline} outlines \companyName{}'s journey to implement the CodeReviewBot, starting with the investigation phase on 25\textsuperscript{th} August 2023, driven by a need to enhance code quality and efficiency. \companyName{} evaluated several open-source projects, including CodeRabbit \cite{CodeRabbit}, Qodo (Formerly CodiumAI)\cite{Codium}, and pipeline extension tools like Reviewbot\footnote{https://github.com/reviewboard/ReviewBot}, ChatGPT-CodeReview\footnote{https://github.com/anc95/ChatGPT-CodeReview}, and Codereview.gpt\footnote{https://github.com/sturdy-dev/codereview.gpt}. After careful consideration, \companyName{} selected Qodo PR-Agent\cite{Codium}, for its high performance and seamless integration. They customized its functionality to meet their needs, and named it "Code Review Bot" internally. For the rest of the study, we will use "CodeReviewBot" to enhance readability. The first pilot project launched on 7\textsuperscript{th}  November 2023. By 5\textsuperscript{th} June 2024, \companyName{} had adopted CodeReviewBot across 10 projects and 22 repositories. On 24\textsuperscript{th} May 2024, practitioners were informed that the commit resolution policy was in place via an email campaign, and 20\textsuperscript{th} September 2024 marked our study's final data collection date. In Figure \ref{fig:CodeReviewBot_Snip} you can see an example comment from CodeReviewBot.

\begin{figure}[h!]
    \centering
    \includegraphics[width=1\linewidth]{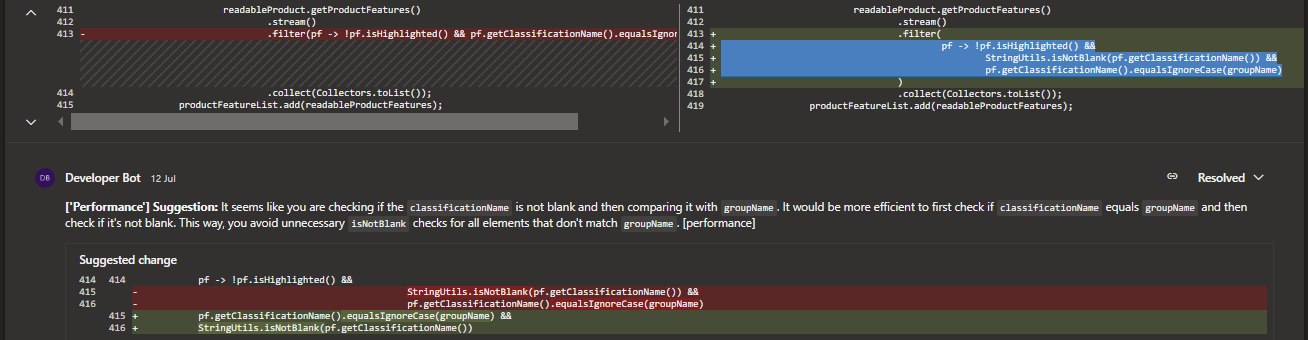}
    \caption{Example Review of CodeReviewBot}
    \label{fig:CodeReviewBot_Snip}
\end{figure}
CodeReviewBot utilizes the GPT-4-32K Model to provide automatic code review comments for each pull request. The tool complements the code review process; developers can still add their reviews. In Figure \ref{fig:Flow_of_PR_PRocess}, the flow of the pull request process is depicted. The tool works on a diverse portfolio of 22 repositories spanning 10 distinct projects. These repositories contained Java, JavaScript, C, HTML, SQL, C\# and TypeScript code. 

\begin{figure}[h!]
    \centering
    \includegraphics[width=1\linewidth]{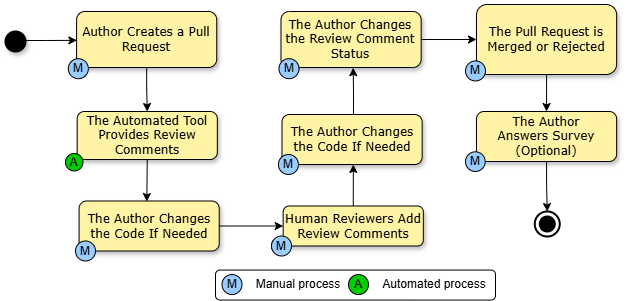}
    \caption{Flow of Pull Request Process}
    \label{fig:Flow_of_PR_PRocess}
\end{figure}

\begin{table}[]
\centering
\caption{Projects Included In Our Study}
\begin{tabular}{|p{1cm}|p{3cm}|p{1.5cm}|p{2cm}|}
\hline
\rowcolor[HTML]{FFFC9E} 
\multicolumn{1}{|c|}{\cellcolor[HTML]{FFFC9E}\textbf{Project}} & \multicolumn{1}{c|}{\cellcolor[HTML]{FFFC9E}\textbf{Explanation}} & \textbf{\# of Developers} & \textbf{Language}                                                                                          \\ \hline
\textbf{\ProjectOne{}}                                         & \begin{tabular}[c]{@{}l@{}}B2B E-commerce \\ Portal\end{tabular}                               & 21                        & \begin{tabular}[c]{@{}l@{}}TypeScript, Java,\\ JavaScript\end{tabular}                                \\ \hline
\textbf{\ProjectThree{}}                                              & \begin{tabular}[c]{@{}l@{}}Enterprise Management \\ Platform\end{tabular}                                   & 51                        & \begin{tabular}[c]{@{}l@{}}HTML, JavaScript,\\ C\#\end{tabular}                                                              \\ \hline
\textbf{\ProjectTwo{}}                                                 & \begin{tabular}[c]{@{}l@{}}Customizable AI \\ Solutions Hub\end{tabular}                                      & 22                        & \begin{tabular}[c]{@{}l@{}}C\#, HTML,\\ JavaScript\\ SQL\end{tabular}                                           \\ \hline

\end{tabular}
\label{fig:Projects_Included}
\end{table}

The \companyName{} was selected as the study object since they had used the CodeReviewBot for a considerable time. The size of their development teams (103 developers) and the diversity of projects (10 projects) employing the tool are other qualities that motivated us to select them. To avoid potential risks or unintended consequences with collecting real-world data, we did not associate any result with an individual practitioner, and the individual survey results were not shared with the authors from \companyName{}. Our data analysis is conducted on three of the ten projects that employed CodeReviewBot. This is due to the early start of these three projects with the CodeReviewBot (Project \#1 7\textsuperscript{th} November 2023, Project \#2 13\textsuperscript{th} November 2023 and Project \#3 3\textsuperscript{rd} April 2024).  The other seven projects adopted CodeReviewBot around June 2024; hence, they did not accumulate considerable data. Table \ref{fig:Projects_Included} provides information regarding the three projects. 

\subsection{Study Goal and Research Questions}
\label{Study-Goal-and-Research-Questions}

Our study investigates the impact of automated code reviews in software development from multiple perspectives. Academically, it provides empirical data to understand the future direction of modern code review and reveals the potential of promising automated code review tools. For practitioners, it is crucial to determine whether automated code reviews positively affect the development process. From a business standpoint, implementing such tools involves costs that their benefits must justify. Quality-wise, automated code reviews should be expected to enhance rather than diminish the existing code review process. Therefore, this study is a well-motivated and valuable investigation for industry professionals and academics.

We addressed the following research questions by collecting data from the project repositories and surveys:

\textbf{Research Question 1:} How useful are LLM-based automated code reviews in the context of the software industry?

This research question assesses LLM-based automated code reviews' utility in software development. Specifically, we seek to determine whether comments generated by automated code review tools are effectively incorporated into code. This depends on the usefulness of the automated reviews and the developer's reception of them. 

\textbf{Research Question 2:} How do LLM-based automated code reviews impact the pace of the pull request closure process?

One of the main motivations for the automation is time gains. With this research question, we examine the effect of LLM-based automated code reviews on the pace of development. Specifically, we aim to determine whether integrating automated code reviews accelerates the pull request closure process. By analyzing this impact, we seek to understand whether LLM-based tools contribute to more efficient development workflows.

\textbf{Research Question 3:} How does the introduction of LLM-based automated code reviews influence the volume of human code review activities? 

Human code reviews require substantial effort. With the introduction of LLM-based automated code reviews, assessing how this affects the volume of human reviews is essential. With this research question, we aim to determine whether automated tools lead to an increase or decrease in human review activities. We aim to understand how automation impacts the effort and workload associated with human code reviews.

\textbf{Research Question 4:} How do developers perceive the LLM-based automated code review tools?

Developers are important actors in the code review process. Their perception of the automated code review tools is critical for realizing the expected benefits. With this research question, we aim to analyze how developers perceive the comments they receive and the general process by triangulating our different data sources. This comprehensive assessment will provide insights into developers’ attitudes toward automation and its role in the code review process. 


We utilized qualitative and quantitative data to achieve our research objectives and address our questions. The interaction between practitioners and automated code reviews is a key focus of our study, which required qualitative data collection. To achieve this, we developed a pull request survey and a general opinion survey and implemented a mandatory comment resolution policy requiring developers to label how comments were addressed. For quantitative data, we opted for repository mining, which offers valuable insights into real-world usage by systematically recording digital activities. We triangulated our findings from the surveys and comment resolution labels by analyzing historical commit data, pull requests, and related comments. Figure \ref{fig:data-collection-overview} depicts our data collection overview.

\subsection{Data Collection From Azure DevOps}
\label{Data-Collection-From-Azure-DevOps}
We collected data from the project's version control system ("Azure DevOps"). This data includes pull request information, review comments, and commits to pull requests. To better understand the data extraction process, we described the components illustrated in Figure \ref{fig:data-collection-overview}.

\begin{figure}[h!]
    \centering   \includegraphics[width=1\linewidth]{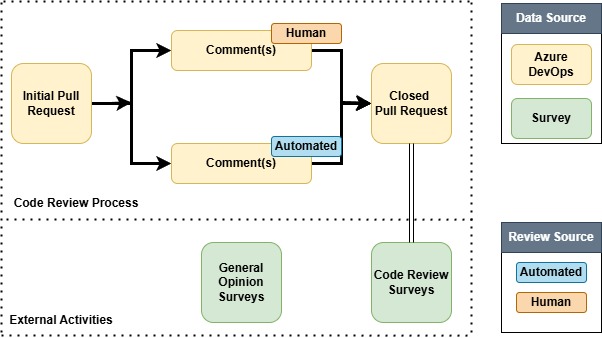}
    \caption{Data Collection Overview}
    \label{fig:data-collection-overview}
\end{figure}

\subsubsection{Initial Pull Request}
A pull request is a mechanism for introducing changes to the code base. Once a pull request is created, other developers are informed about the proposed changes. They add their review comments to notify the author, who is expected to make necessary changes based on the reviews. The pull request is accepted once the code reaches acceptable quality and the changes are merged into the code base. 

Pull request authors receive automatic review comments and other reviewers' comments in our case. Our study considers \totalAnalyzedPR{} pull requests, of which \totalAnalyzedPRAfterCodium{} were created after adopting automated code review.



\subsubsection{Pull Request Comments}
Pull request comments serve as the artifacts of the code review process. Human reviewers add their reviews in the form of comments. The CodeReviewBot added its reviews as separate comments. We analyzed the pull request comments regarding volume and time, who generated them, and how pull request authors benefited. 

The presence of a comment does not mean it was taken into consideration. To investigate whether pull request authors benefited from the review comments, we configured a required comment resolution policy \cite{Azure_Comment_Resolution}. This system requires authors to indicate how they handled the pull request comments. Unfortunately, the comment resolution policy put in place after CodeReviewBot, so we do not have the same data from prior pull requests. Table \ref{comment-status-options} presents the comment status options for each review comment and their explanation according to the comment resolution policy.

\begin{table}[h!]
\centering
\caption{Comment Resolution Policy in Azure DevOps}
\resizebox{0.5\textwidth}{!}{
\begin{tabular}{|l|l|}
\hline
\rowcolor[HTML]{FFFC9E} 
\textbf{Status}                           & \textbf{Explanation}                                           \\ \hline
{\color[HTML]{222222} \textbf{Active}}    & {\color[HTML]{222222} This is the default status for new comments.}                                                                           \\ \hline
{\color[HTML]{222222} \textbf{Pending}}   & {\color[HTML]{222222} \begin{tabular}[c]{@{}l@{}}The corresponding comment is being analyzed\\ or waiting for some other thing.\end{tabular}} \\ \hline
{\color[HTML]{222222} \textbf{Resolved}}  & {\color[HTML]{222222} \begin{tabular}[c]{@{}l@{}}The corresponding comment is successful and\\ its suggestion was implemented.\end{tabular}}  \\ \hline
{\color[HTML]{222222} \textbf{Won’t fix}} & {\color[HTML]{222222} \begin{tabular}[c]{@{}l@{}}The corresponding comment is faulty or cannot\\ be implemented.\end{tabular}}                \\ \hline
{\color[HTML]{222222} \textbf{Closed}}    & {\color[HTML]{222222} \begin{tabular}[c]{@{}l@{}}The corresponding comment was not implemented\\ for some other reasons than "Won't Fix".\end{tabular}}        \\ \hline
\end{tabular}
}
\vspace{0.2cm}
\label{comment-status-options}
\end{table}



\subsubsection{Closed Pull Request}
The code changes proposed in the initial pull request may change with the review comments. This is an expected event if the reviews point out problems. The code changes are reflected as additional commits. The pull request is either accepted or rejected. The practitioners at \companyName{} are not using rejection as a quality assurance mechanism. Instead, they use rejection when the change is considered unnecessary or untimely. Hence, we did not treat acceptance as a quality indicator but as categorical data. 



\subsubsection{Data Extraction and Analysis}


Our data analysis starts with extracting raw data from the Azure DevOps server using the API. We created a data scheme that allowed us to store information regarding projects, repositories, commits, pull requests, and pull request comments in a relational database. We used the data scheme to store the API responses. 

Our next step was to preprocess the data to ensure its integrity and accuracy. Firstly, we converted the tables into CSV files, uploaded the data using the pandas library\footnote{https://pandas.pydata.org/pandas-docs/stable/index.html\#}, and created scripts to extract the relevant metrics. The first problem we noticed was the high number of active comments, which should not have been the case for the comment resolution policy. We manually examined pull requests and observed that some pull requests were closed before the CodeReviewBot could comment. The comments created by the tool arrived after the pull request was closed, meaning they did not need to be resolved. 
We reviewed the time it took to merge the PRs and, using elbow evaluation, identified a threshold that covers 93\% of the PRs. We then applied this filter to the entire history for a fair evaluation. Finally, we manually removed the remaining 7\% to eliminate all outliers. 

The second issue we encountered was the high number of comments from some developers. Two developers had unreasonably higher comments than other developers. We examined their comments and realized that their Azure DevOps accounts were used for different software bots, such as SonarQube\footnote{https://www.sonarsource.com/products/sonarqube/}. We also excluded these accounts from our study. 

Lastly, we observed that many comments had unknown comment resolution labels due to being deleted or including the pull request description generated by CodeReviewBot. For that, we excluded such comments. On top of that, the comment resolution policy was only active when the pull request targeted the repository's main branch. For that, we excluded pull requests targeting other branches except for the main branch. The data cleaning process involved several collaborative sessions with both non-practitioner and \companyName{} participants, after which we confirmed the dataset was free from corruption. Ultimately, we analyzed 4,335 pull requests, of which 1,568 were subject to automated reviews.

\subsection{Data Collection From Surveys}
\label{Data-Collection-From-Surveys-and-Interviews}
\subsubsection{Code Review Surveys}
With each pull request, the authors were asked to give a zero to five rating for the automated review comments and received a survey of three questions. These surveys are our second data source. The first two questions have one-to-five scale answers, and the last one is open-ended. We included code review survey questions in the
replication package.

\subsubsection{General Opinion Surveys}

We created a general opinion survey and received responses from \surveyRespondents{} developers who contributed to the three projects within our research scope. These practitioners had different years of experience and positions in the organization. We provide information regarding the participants in Table \ref{tab:survey-participants}. The survey questions focused on the impact of automated code reviews on developers and their perspectives. We included general opinion survey questions in the replication package. 

\begin{table}[h!]
\centering
\caption{Survey Participants}
\begin{tabular}{|l|r|}
\hline
\multicolumn{2}{|c|}{\cellcolor{blue!80}\textcolor{white}{\textbf{Survey Participants}}} \\
\hline
\rowcolor{blue!10}
\textbf{Experience in Software Development} & \textbf{Number of Participants} \\
\hline
0-2 & 4\\
\hline
2-5 & 9 \\
\hline
5-10 & 6 \\
\hline
10+ & 3 \\
\hline
\rowcolor{blue!10}
{\textbf{Position at \companyName{}}}   & \textbf{Number of Participants} \\
\hline
Individual Contributor & 16 \\
\hline
Lead/Manager & 6 \\
\hline
\rowcolor{blue!10}
\textbf{Total Practitioners} & \textbf{22}\\
\hline
\end{tabular}
\label{tab:survey-participants}
\end{table}



\subsection{Approach Towards Research Questions}
\label{Approach-Towards-Research-Questions}
Since our research questions have multiple aspects, we relied on findings from multiple data sources. We used the review comment labels, commits to pull requests after reviews, and general opinion survey question answers to establish a multi-aspect evaluation for the first research question. We extracted pull request closure duration data from Azure DevOps to answer the second research question. In the general opinion survey, we asked the practitioners whether automated code reviews affected the development pace. 

We extracted the number of human code reviews from Azure DevOps for the third research question. We asked developers whether they could manually generate the same comments in the survey. The last research question is centered around the developer and pull request survey. Our data analysis scripts, survey questions, and results are shared in our replication package\footnote{https://doi.org/10.5281/zenodo.13917481}.

\section{Results}
\newcommand{\ProjectOne}{Project \#1}
\newcommand{\ProjectTwo}{Project \#3}
\newcommand{\ProjectThree}{Project \#2}
\label{results}
\subsection{Code Review Surveys}

During our study, we collected ratings for 38 pull requests that received automated code reviews. The pull request authors were also asked to fill out a survey consisting of two multiple-choice questions and one open-ended question, and ten people went on to do so.  

The average rating for the automated review comments was 3.46, with a standard deviation of 1.79. Figure \ref{fig:code_review_ratings} depicts the ratings for the different projects and the overall rating. There is a considerable difference between ratings across projects, with averages of 4.04, 2.72, and 3.00 for  \ProjectOne, \ProjectThree, and \ProjectTwo,  respectively. This might indicate a difference in the quality of reviews concerning different project conditions, or it may be due to the differences between developers in different projects. 

The survey that we sent to the authors contained three questions. The first two questions were regarding whether authors found the reviews agreeable and how they found the presentation of the review comments. The third question was open-ended and asked for additional feedback. 

\begin{figure}[ht]
  \centering   \includegraphics[width=1\linewidth]{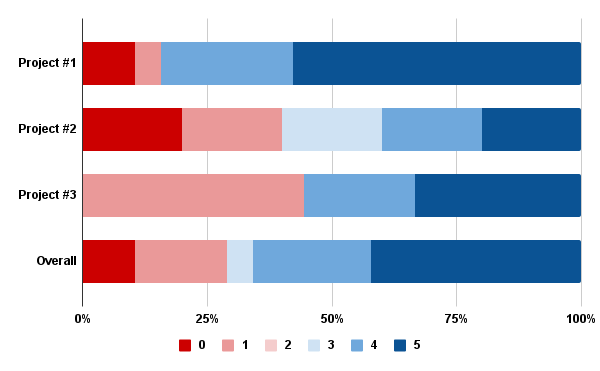}
    \caption{Automated Code Review Ratings from Code Review Surveys}
  \label{fig:code_review_ratings}
\end{figure}

Unfortunately, we did not receive many answers for the third question. Most of the answers were simple remarks such as "Great." Ten practitioners answered the first two questions, and we reported on the responses in Figure \ref{fig:pull_request_survey}. Eight people answered with "5" when asked how agreeable they found the comments, while nine answered with "5" when asked how well-presented they found the comments. One person answered both questions with "1". There is a difference between the number of responses for ratings and surveys. Since the survey required extra effort, dissatisfied developers might have felt less motivated to complete the survey. Overall, the ratings and the pull request survey show developers found the comments agreeable and well-presented.

\begin{figure}[ht]
\centering   \includegraphics[width=1\linewidth]{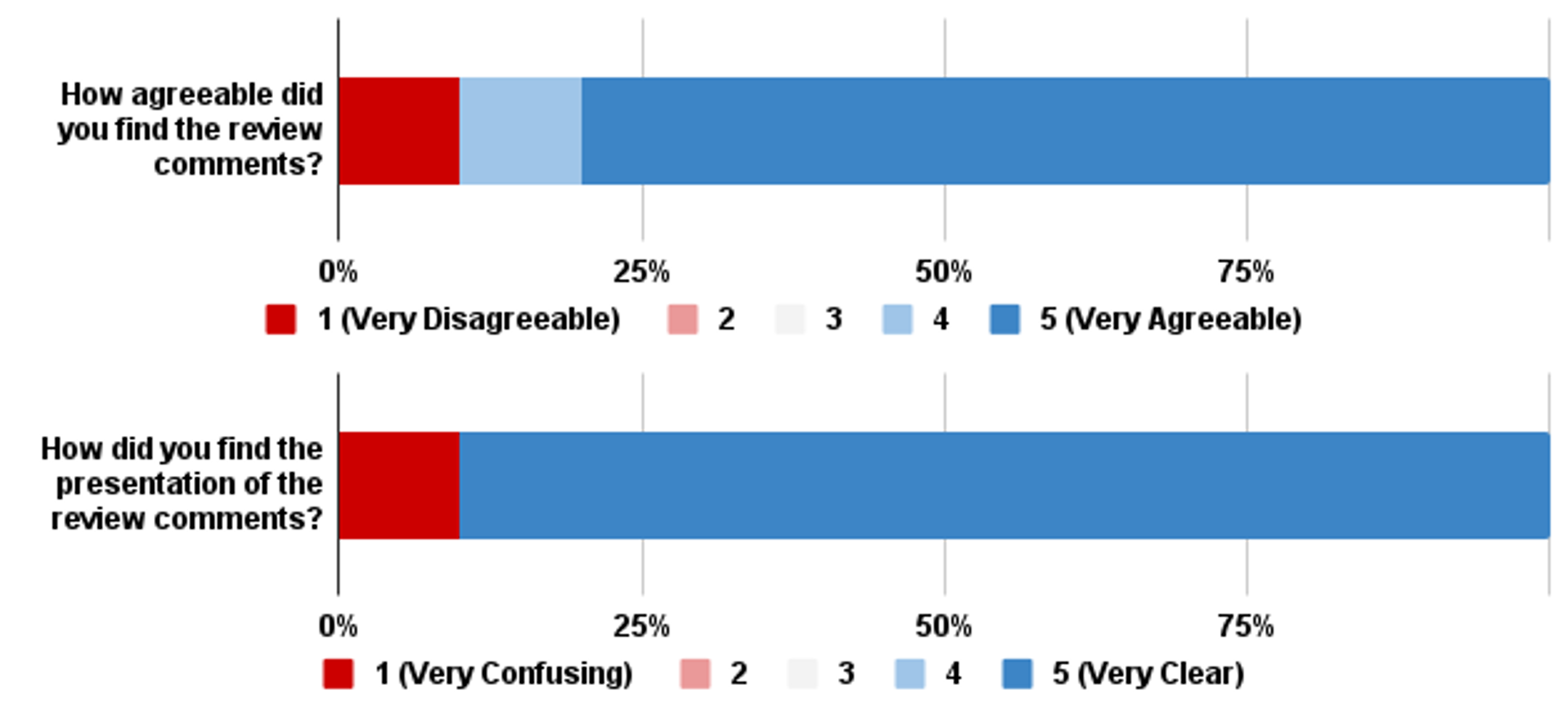}
  \caption{Code Review Survey}
  \label{fig:pull_request_survey}
\end{figure}

\subsection{General Opinion Surveys}
We sent a survey consisting of eight questions to the developers who contributed to the projects. We received 23 responses, with 22 respondents consenting to be involved in published results. This survey was created to collect the overall perceptions of developers.

The survey comprised six multiple-choice questions and two open-ended questions. Figure \ref{fig:first_three_questions} and \ref{fig:likert_scale_survey} present the results. The first question addressed the impact of automated code reviews on development speed. Eight developers perceived a minor improvement in the pace of development due to automated code reviews, while four perceived no effect, three perceived a minor deterioration, and two perceived a major improvement.  

\begin{figure}[ht]
\centering   \includegraphics[width=1\linewidth]{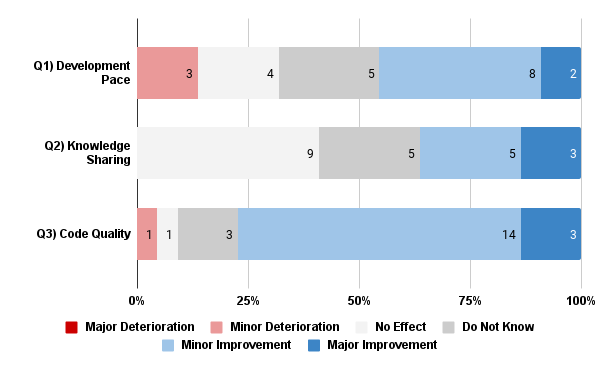}
  \caption{General Opinion Survey Responses (Questions 1-3)}
  \label{fig:first_three_questions}
\end{figure}

The second question focused on knowledge sharing. Nine practitioners perceived no effect on knowledge sharing among developers. At the same time, the eight developers perceived a positive impact. The third question explored code quality, with most respondents (14) suggesting that automated code reviews contribute to a minor improvement in the overall quality of the code.

\begin{figure}[ht]
\centering   \includegraphics[width=1\linewidth]{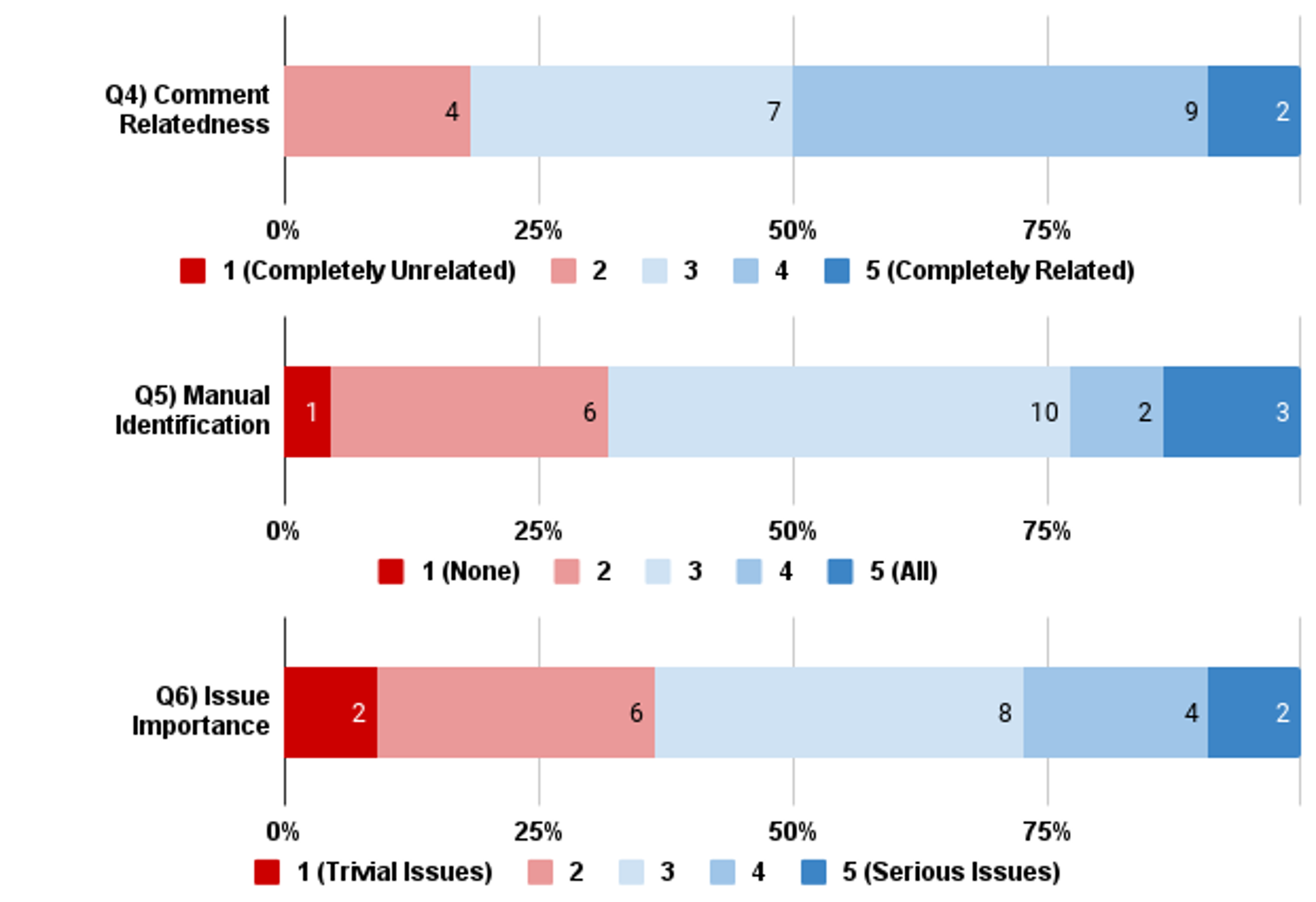}
  \caption{General Opinion Survey Responses (Questions 4-6)}
  \label{fig:likert_scale_survey}
\end{figure}

The fourth question examined the relevance of the comments generated by the automated code reviews to the pull request. With nine respondents rating the relatedness as "4" and seven rating the relatedness as "3", results indicate that most practitioners found the automated comments related to pull requests.

The fifth question asked whether developers could manually identify the issues the automated code reviews pointed out with "1," meaning they could not manually identify none, and "5," meaning they could identify all of them. Ten respondents rated this question "3," while six rated it "2". 

The sixth question addressed the importance of the issues highlighted by the automated reviews. Eight respondents rated this question "3," while six rated it "2". The dispersion of the ratings shows that practitioners have different views.

\subsection{Analysis of the Azure DevOps Data}
\newcommand{\wontFixTotalPercent}{21.3\%}
\newcommand{\wontFixTotalNumber}{300}
\newcommand{\ResolvedTotalPercent}{73.8\%}
\newcommand{\ResolvedTotalNumber}{1040}
\newcommand{\ClosedTotalPercent}{4.8\%}
\newcommand{\ClosedTotalNumber}{68}
\newcommand{\TotalCommentByBot}{4408}
\newcommand{\TotalCommentByBotAnalyzed}{1408}
\newcommand{\TotalPullRequests}{4335}
\newcommand{\TotalPullRequestsAnalyzed}{6000}
\newcommand{\PullRequestClosureProjectOneBefore}{two hours and 48 minutes}
\newcommand{\PullRequestClosureProjectTwoBefore}{20 hours and 22 minutes}
\newcommand{\PullRequestClosureProjectThreeBefore}{six hours and six minutes}
\newcommand{\PullRequestClosureOveralBefore}{five hours and 52 minutes}
\newcommand{\PullRequestClosureProjectOneAfter}{four hours and 38 minutes}
\newcommand{\PullRequestClosureProjectTwoAfter}{30 hours and 51 minutes}
\newcommand{\PullRequestClosureProjectThreeAfter}{three hours and seven minutes} 
\newcommand{\PullRequestClosureOveralAfter}{eight hours and 20 minutes}
\newcommand{\TimeLimitForPullRequest}{5}
\newcommand{\AvgBotComments}{4}

\begin{figure}[ht]
\centering   \includegraphics[width=1\linewidth]{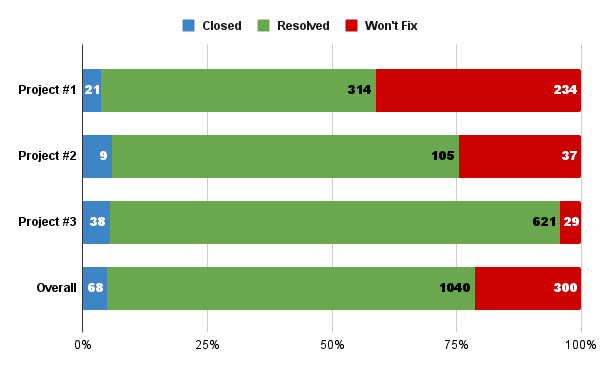}
    \caption{Comment Labels on Merged PRs}
\label{fig:comment_labels}
\end{figure}

\subsubsection{Comment Labels}
Within our analysis, we extracted \TotalCommentByBot{} comments by CodeReviewBot. Since the comment resolution policy was enforced sometime after the CodeReviewBot, we had to filter out comments before the policy introduction and the currently active or pending comments. After the filtering, we examined \TotalCommentByBotAnalyzed{} CodeReviewBot comments on merged pull requests. We reported the status of comments for each project in Figure \ref{fig:comment_labels}. \ResolvedTotalPercent{} of the comments are labeled as "Resolved," while \wontFixTotalPercent{} are labeled as "Won't Fix." When we compare the results across projects, we see a significant difference for \ProjectOne{} and \ProjectTwo{} with 55\% and 90\% of comments labeled as "Resolved."
\begin{figure}[!ht]
\centering   \includegraphics[width=1\linewidth]{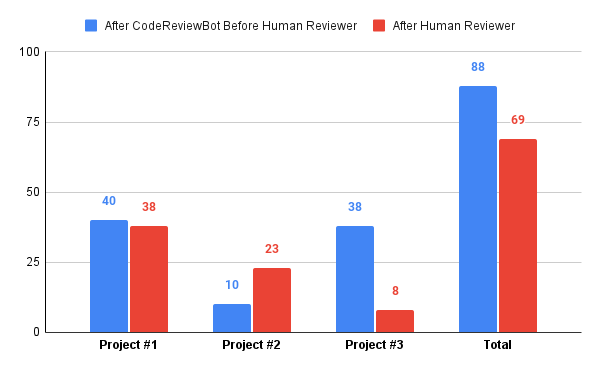}
  \caption{Commits After Code Reviews}
\label{fig:commits_after}
\end{figure}

\subsubsection{Commits on Pull Requests}
As a measure of change caused by the code review process, pull requests receive commits after they receive comments. We analyzed these commits for the pull requests after CodeReviewBot's implementation. The results are shown in Figure \ref{fig:commits_after}. Pull requests received 88 commits after CodeReviewBot's comments but before human comments. After the human reviewer had commented, the pull requests received 69 commits.
It should be noted that authors may also act upon CodeReviewBot's comments after the human reviewer's comments. Conversely, some pull requests may have been opened as drafts, with new commits added afterward. A difference between the number of commits after CodeReviewBot and the comments with the "Resolved" label is expected. Since pull requests receive more than one comment and commits are made for the pull request, the "Resolved" labeled comments should be more than the commits. However, developers might also disregard the expected labeling policy and use "Resolved" instead of "Closed" or "Won't Fix." "Closed" refers to comments that were not faulty but were not implemented for some other reason. 
%

\begin{tcolorbox}[colback=black!5!white, colframe=black!75!black, title=RQ1: How useful are LLM-based automated code reviews in the context of the software industry?]
Our analysis showed that \ResolvedTotalPercent{} of the comments suggested by CodeReviewBot were accepted and implemented by developers,  highlighting the bot's significant impact on the code review process. Moreover, 88 commits were made after CodeReviewBot and before human reviewers' comments, indicating proactive changes based on the bot's suggestions. Most survey respondents also perceived an improvement in code quality using CodeReviewBot. Therefore, we conclude that LLM-based automated code reviews are highly useful in this company's context.
\end{tcolorbox}

\subsubsection{Pull Request Closure Durations}
\begin{figure}[ht]
\centering   \includegraphics[width=1\linewidth]{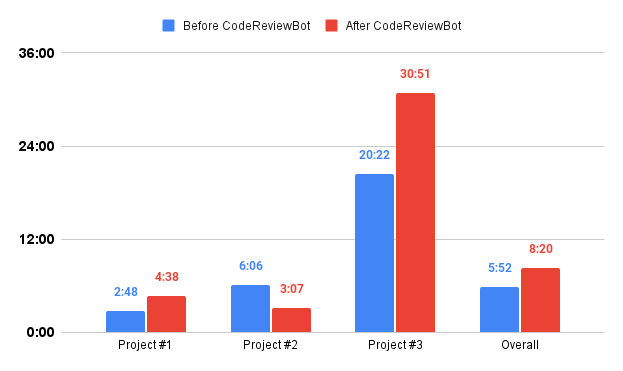}
  \caption{Pull Request Closure Durations Before and After CodeReviewBot (H:MM)}
  \label{fig:pull_request_closure}
\end{figure}
The overall average pull request closure duration increased from five hours and 52 minutes before the introduction of CodeReviewBot to eight hours and 20 minutes after its implementation. The independent samples t-test indicated this overall increase was statistically significant (p-value $<$ 0.001). We observed different trends across projects as depicted in Figure \ref{fig:pull_request_closure}.

Project \#1 showed a significant increase in closure duration, rising from two hours and 48 minutes to four hours and 38 minutes after CodeReviewBot was introduced. The t-test indicated this increase was statistically significant (p-value $<$ 0.001). Project \#2 experienced a significant decrease in closure duration, dropping from six hours and six minutes to three hours and seven minutes. The analysis yielded a statistically significant result (p-value $<$ 0.001). Project \#3 observed an increase in closure duration from 20 hours and 22 minutes to 30 hours and 51 minutes after the bot's introduction. The statistical test showed this increase was statistically significant (p-value $<$ 0.001).

\begin{tcolorbox}[colback=black!5!white, colframe=black!75!black, title=RQ2: How do LLM-based automated code reviews impact the pace of the pull request closure process? ]
Our results indicated a significant slowdown in the pull request closure process. This slowdown could be explained by developers needing to address additional comments from the bot and those from human reviewers. The impact on closure durations varied across different projects, suggesting that project-specific conditions play a critical role in determining the effects.
\end{tcolorbox}

\subsubsection{Human Reviewer Comments}

\begin{figure}[!ht]
\centering   \includegraphics[width=1\linewidth]{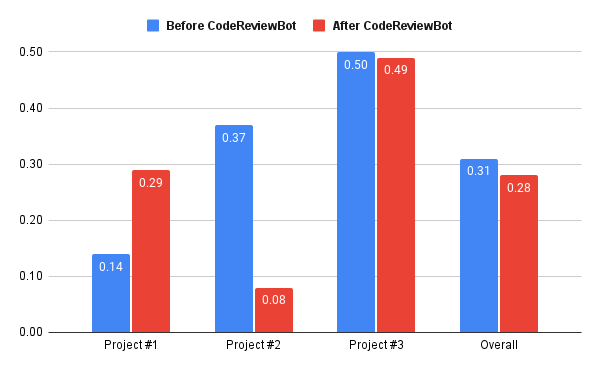}
    \caption{Average Number of Human Reviewer Comments}
\label{fig:human_reviwer_comments}
\end{figure}

Overall, human reviewers left an average of 0.31 comments per pull request before deploying CodeReviewBot, which decreased to 0.28 afterward, as shown in Figure \ref{fig:human_reviwer_comments}. The Poisson regression analysis indicated that this overall decrease was not statistically significant (p-value $\geq$ 0.05). For comparison, CodeReviewBot left a higher average of 3.65 comments per pull request. The statistical analysis revealed different trends across projects:

After introducing CodeReviewBot, human comments per pull request significantly increased in Project \#1 (from 0.14 to 0.29, p-value $<$ 0.05), significantly decreased in Project \#2 (from 0.37 to 0.08, p-value $<$ 0.05), and remained statistically unchanged in Project \#3 (from 0.50 to 0.49, p-value $\geq$ 0.05).

\begin{tcolorbox}[colback=black!5!white, colframe=black!75!black, title=RQ3: How does the introduction of LLM-based automated code reviews influence the volume of human code review activities?]
The average number of human comments decreased from 0.31 to 0.28 with the introduction of CodeReviewBot. However, according to our Poisson regression analysis, this decrease was not statistically significant.  Additionally, the impact of LLM-based automated code reviews on human reviewer activity varied across different projects, potentially influenced by project dynamics and team practices.
\end{tcolorbox}


\subsection{General Opinion Survey Open-Ended Questions}

The general opinion survey had two questions about the advantages and disadvantages of automated code review. We received 20 legitimate answers, with two non-useful responses (blank spaces). A significant proportion of participants (13/20) remarked on advantages regarding code quality improvement and the maintenance of coding standards. Some respondents highlighted the enhancements in the review process, such as shortening the review process and helping spot overlooked code smells and potential bugs. The auto-generated code descriptions were cited as beneficial in expediting the review process. Other notable advantages included providing suggestions for improvement and enhancing awareness of best practices.

\begin{tcolorbox}[colback=black!5!white, colframe=black!75!black, title=RQ4: How do developers perceive the LLM-based automated code review tools?]
Some practitioners voiced concerns that out-of-scope or irrelevant suggestions could slow down reviews and create distractions. Some practitioners were concerned that automated code reviews missed critical issues that a human reviewer would catch. One respondent raised a significant concern about automated code review potentially altering the context of pull requests, which could introduce severe bugs if changes are applied without careful consideration. Despite these drawbacks, the majority of developers perceive automated code review tools as beneficial for improving code quality and maintaining coding standards, notably by detecting quality problems and providing improvement suggestions.
\end{tcolorbox}

\section{Discussion}
\label{discussion}
\subsection{Does LLM-based automated code review considerably improve software development activity?}
\newcommand{\commentLabelResolvedPercentage}{73.8\%}
\newcommand{\commentLabelWontFixClosedPercentage}{26.2\%}
\newcommand{\commitAfterReview}{54\%}
\newcommand{\minorImprovementCodeQuality}{68.8\%}
\newcommand{\minorImprovementDevelopmentPace}{50\%}
\newcommand{\averagePRClosureBefore}{50 minutes}
\newcommand{\averagePRClosureAfter}{48 minutes}
\newcommand{\averageImportance}{2.78}
  
Implementing an LLM-based automated code review tool is a substantial organizational decision that involves both benefits and costs. While such tools incur expenses—for instance, in our study, the CodeReviewBot used an average of 3,937 tokens per pull request at a cost of  0.48\$ — they also require developers to invest time in addressing review comments.

The comment labels assigned by pull request authors showed that \commentLabelResolvedPercentage{} of the comments were addressed by the author. Additionally, we observed that 88 commits were made after the CodeReviewBot's reviews before human reviews, signifying that the automated reviews affected the pull requests. Our survey of developers revealed that \minorImprovementCodeQuality{} perceived a minor improvement in the code quality after CodeReviewBot. According to our survey results, practitioners consider issues pointed out by CodeReviewBot important and related to the respective pull requests. This further supports the usefulness of LLM-based automated code reviews.

On top of usefulness for code quality, code reviews are also helpful for knowledge sharing \cite{Macleod2018_CRInTrenches}. Within our general opinion survey, we asked practitioners whether CodeReviewBot affected knowledge-sharing. Most respondents cited no effect, while no respondent pointed to a negative impact. 

One of the cited benefits of automated code review is its potential to save time and developer effort \cite{Tufano2021_TowardsAutomatingCR}. To explore this, we analyzed the durations for pull request closures. Although we observed an overall increase in closure times, this could be attributed to authors spending extra time fixing issues highlighted by the automated reviewer bot. The trends varied significantly across different projects, suggesting that developers were actively engaging with the automated feedback, which may have extended the closure times but potentially led to higher code quality. Furthermore, our analysis showed that the number of human review comments per pull request did not decrease significantly after introducing the automated tool. This suggests that while developers invested additional effort to address automated comments, it did not replace the need for human reviews. Consequently, we did not find conclusive evidence supporting consistent time or effort savings as a result of implementing the automated code review tool.

Our findings suggest that automated code review can moderately improve software development activity. The decision to implement such a tool should still be considered carefully. The observed benefits within our study might be hindered by existing code review habits or other quality assurance activities in place. 

\subsection{Implications to Practitioners}

\textbf{Over-reliance on automated code reviews}: One respondent in the general opinion survey said, "It may create bias so reviewers may ignore by saying that if any other issue exists, the bot would have written it." This over-reliance on automation can be harmful to the organization by allowing severe bugs to go unnoticed. Practitioners should examine LLM-based automated code review tools extensively before widespread adoption. Organizational awareness of the limitations should be established, and necessary precautions should be taken.  

\textbf{Unnecessary Review Comments}: The results from the comment resolution labels show that \commentLabelWontFixClosedPercentage{} of the tool comments were not acted upon as they were labeled with "Won't Fix" or "Closed". These comments could be too trivial, unrelated or not a problem for the context of the pull request. In any way, developers spend time on them. The survey respondents also cited this problem. One respondent in the general opinion survey said, "It also makes suggestions that fix code blocks that are not in the scope of the task," and another said, "Sometimes the mistakes it thinks it finds are not mistakes at all."

\textbf{Early Identification of Bugs:} The survey showed that early identification of bugs is an advantage of the automated code review. One respondent in the general opinion survey said "It makes finding code defects more easy. Developers can see their mistakes fast." Respondents also highlighted the tool's ability to detect typos and forgotten test code.  Another said, "very effective in detecting overlooked code smells". These remarks show us how the tool can improve code quality.


\subsection{Implications to Researchers}
\textbf{Effectiveness of LLM-Based Tools}: This study provides valuable empirical evidence on the effectiveness of LLM-based automated code review tools in real-world industry settings. The positive reception of developers and the percentage of comments accounted for in the pull requests (\commentLabelResolvedPercentage) suggest that LLMs can enhance the code review process. One respondent in the general opinion survey said, "It improved the awareness of the team about code quality". This remark suggests that the tool also has effects beyond code review practice.


\textbf{Human-AI Interaction Dynamics}: Integrating automated reviews into the code review process introduces new dynamics in human-AI interaction. One of the findings was the frustration caused by recursive reviews, with one respondent saying, "With each fix, a new review is generated. However, after the initial review, subsequent comments on revised pull requests often become redundant and unhelpful.". Researchers can investigate factors influencing developer trust, satisfaction, and reliance on automated reviews, leading to more human-centered design improvements in these tools.


\section{Threats to Validity}
\label{threats_to_validty}
\subsection{Construct Validity}
Our data collection process involves collecting data from respondents through survey answers and comment resolution labels. These data entries are subject to misinterpretation. To mitigate this threat to validity, extra effort was put into explaining the expectations from the respondents. We also thoroughly reviewed the questions with multiple reviewers to rephrase potentially confusing and complicated language.

\subsection{Internal Validity}
Some authors of this paper are part of Beko's software development organization. Specifically, one of the authors is in a managerial position in the organization. This introduces the possibility of biases. For having a neutral stance, the results and discussions in this paper were created by the non-practitioner authors. 

We acknowledge that the mandatory comment resolution policy could be considered time-consuming for practitioners because they do not provide reliable comment resolution labels. To account for this threat, we triangulated our data sources to come to conclusions. For example, we used comment labels alongside pull request change information to see whether they agreed to a certain extent.


Our data collection is centered around summer. Beko practitioners mentioned that more developers are on vacation during the summer, which decreases productivity and development pace. For this matter, we acknowledge that our conclusions are subject to seasonality.


The projects examined in this study are real projects that commenced well before the research began. We did not associate any results with individual practitioners to avoid unintended effects and ethical concerns. As a result, the practitioners involved in the projects had no incentive to behave differently since the analysis was conducted independently afterward.


Since the pull request authors the same pull request survey repeatedly, we consider that they might become frustrated and respond carelessly. To avoid this decline in engagement, we decided to limit the survey to three questions. 

Our quantitative data analysis relies on the integrity of data in Azure DevOps. Corruptions in the database could impact our results. To mitigate such corruption, we conducted joint data cleaning sessions where we discussed the quality of the data. We identified two accounts involved in the projects that were used as bot accounts for a certain time. We removed the comments made by those accounts. We acknowledge that those comments might include human comments, though most were from bots.

Although the "Active" and "Pending" comments would not allow the pull request to be closed, we observed that some automated comments arrived after the pull request was closed. These comments were mostly left as active, and we disregarded them in our analysis.

\subsection{External Validity}

Our study was conducted with a certain automated code review tool (Qodo PR Agent\cite{Codium}) based on a certain LLM (GPT-4 32k\cite{Achiam2023_GPT4}). Since we used this specific tool and model, we acknowledge that other LLMs and automated code review tools might exhibit different behavior. Therefore, further research is needed to determine if other LLMs would behave in a comparable manner. 

\subsection{Conclusion Validity}
This study might have led to different results in a different company setting. We cannot account for the many changing conditions to reach statistical conclusions; therefore, we considered a case study the most appropriate approach. A multiple case study examination using our methodology would be required to reach definitive conclusions. Given the challenges of conducting such a comprehensive study, we limited our scope and did not aim for statistical generalizations.

\section{Conclusion}
\label{conclusion}
\newcommand{\commentLabelResolvedPercentage}{73.8\%}
\newcommand{\commitAfterReview}{\%54 }
\newcommand{\minorImprovementCodeQuality}{68.8\% }
\newcommand{\minorImprovementDevelopmentPace}{50\% }
\newcommand{\averagePRClosureBefore}{five hours and 52 minutes}
\newcommand{\averagePRClosureAfter}{eight hours and 20 minutes}
\newcommand{\pointoutmanually}{}

In this study, we conducted an evaluative case study within the software development division of Beko, a multinational company, focusing on automated code reviews. An LLM-based automated code review tool, based on the open-source Qodo PR-Agent \cite{Codium}, was adopted across ten projects. We narrowed our research scope to three of these projects, selected based on their longer duration of tool usage.


Our findings suggest that the tool was effective, with \commentLabelResolvedPercentage{} of the comments being acted upon. Additionally, most developers reported a minor improvement in code quality in the survey and no deterioration in knowledge sharing. Regarding pull request closure times, there was a statistically significant increase between the average times of \averagePRClosureBefore{} before and \averagePRClosureAfter{} after the tool's introduction. Across projects, there were different trends, with one project's pull request closure duration declining. Additionally, there was no significant change in the number of human code reviews before and after the tool's implementation.

Since developers are the primary users interacting with the tool, we examined their perceptions of automated code reviews. The survey revealed that practitioners commonly perceived the issues identified in automated reviews as important and relevant to the pull requests. Participants noted several advantages contributing to code quality, including faster bug detection, eliminating code smells, increased awareness of code quality, and the promotion of standardized best practices. The disadvantages included suggestions for out-of-scope code changes and occasional misaligned recommendations that distract developers and slow down the development. There were also concerns about the downsides of potential over-reliance on automated systems.

Our study revealed that automated code reviews can positively impact software development; however, some unintended effects and disadvantages were also identified. Practitioners can use these insights to make more informed decisions on the implementation and use of LLM-based automated code review tools. We aim to replicate our study with different software development companies to account for inter-organizational differences in our future work.

\section{Acknowledgements}
\label{Acknowledgements}
This work has been supported the ITEA4 GENIUS project, which has been funded by
the national funding authorities of the participating countries:
https://itea4.org/project/genius.html

\bibliographystyle{IEEEtran}
\bibliography{IEEEabrv,sections/references}

\end{document}